\documentclass[twocolumn,showpacs,superscriptaddress,floatfix]{revtex4}%
\usepackage{graphicx}
\usepackage{times,bm}
\usepackage{amsmath,amsfonts,amssymb}

\allowdisplaybreaks[4]

\begin{document}
\title{
Spin-triplet superconductivity in Sr$_2$RuO$_4$ due to 
orbital and spin fluctuations: 
Analyses by \\ 
two-dimensional renormalization group theory
and self-consistent vertex-correction method
}

\author{Masahisa Tsuchiizu}
\author{Youichi Yamakawa}
\address{Department of Physics, Nagoya University,
Furo-cho, Nagoya 464-8602, Japan}

\author{Seiichiro Onari}
\address{Department of Physics, Okayama University,
Okayama 700-8530, Japan}

\author{Yusuke Ohno}
\author{Hiroshi Kontani}
\address{Department of Physics, Nagoya University,
Furo-cho, Nagoya 464-8602, Japan}

\date{April 9, 2015}

\begin{abstract}
We study the mechanism of the triplet superconductivity (TSC)
in Sr$_2$RuO$_4$ based on the multiorbital Hubbard model.
The electronic states are studied 
using the recently developed renormalization group  method
combined with the constrained random-phase-approximation,
called the RG+cRPA method.
Thanks to the vertex correction (VC) for the susceptibility,
which is dropped in the mean-field-level approximations,
strong orbital and spin fluctuations at ${\bm Q}\approx(2\pi/3,2\pi/3)$
emerge in the quasi one-dimensional Fermi surfaces (FSs)
composed of $d_{xz}+d_{yz}$ orbitals.
Due to the cooperation of both fluctuations,
we obtain the triplet superconductivity in the $E_u$ representation,
in which the superconducting gap is given by the 
linear combination of 
$(\Delta_x({\bm k}),\Delta_y({\bm k}))\sim (\sin 3k_x,\sin 3k_y)$.
Very similar results are obtained by applying the diagrammatic calculation
called the self-consistent VC method.
Thus, the idea of ``orbital+spin fluctuation mediated TSC'' 
is confirmed by both RG+cRPA method and the self-consistent VC method.
We also reveal that a substantial superconducting gap 
on the $d_{xy}$-orbital FS
is induced from the gaps on the quasi one-dimensional FSs,
in consequence of the large orbital-mixture 
due to the 4$d$ spin-orbit interaction.

\end{abstract}



\maketitle

\section{Introduction}
\label{sec:Intro}

\vspace*{-.2cm}

Sr$_2$RuO$_4$ is an unconventional superconductor 
with the transition temperature $T_{\rm c}=1.5$K 
\cite{Maeno,Maeno2,Sigrist-Rev}.
This material has been attracting great attention 
since the spin triplet superconductivity (TSC) is indicated by 
the NMR study \cite{Ishida}.
The chiral $p$-wave ($p_x+ip_y$) TSC,
which is analogous of the A-phase of the superfluid $^3$He,
had been predicted in Ref.\ \cite{Rice}.
However, in contrast to superfluid $^3$He,
the paramagnon mechanism is not realized in Sr$_2$RuO$_4$ since
no ferro-magnetic fluctuations are observed.
Instead, strong antiferro-magnetic (AFM) fluctuations with 
${\bm Q}\approx (2\pi/3,2\pi/3)$
are observed by neutron scattering spectroscopy \cite{neutron}.
Since the AFM fluctuations usually drive the 
spin singlet superconductivity (SSC),
the mechanism of the TSC in Sr$_2$RuO$_4$
has been a long-standing problem in strongly correlated electron systems.

\begin{figure}[t]
\includegraphics[width=.9\linewidth]{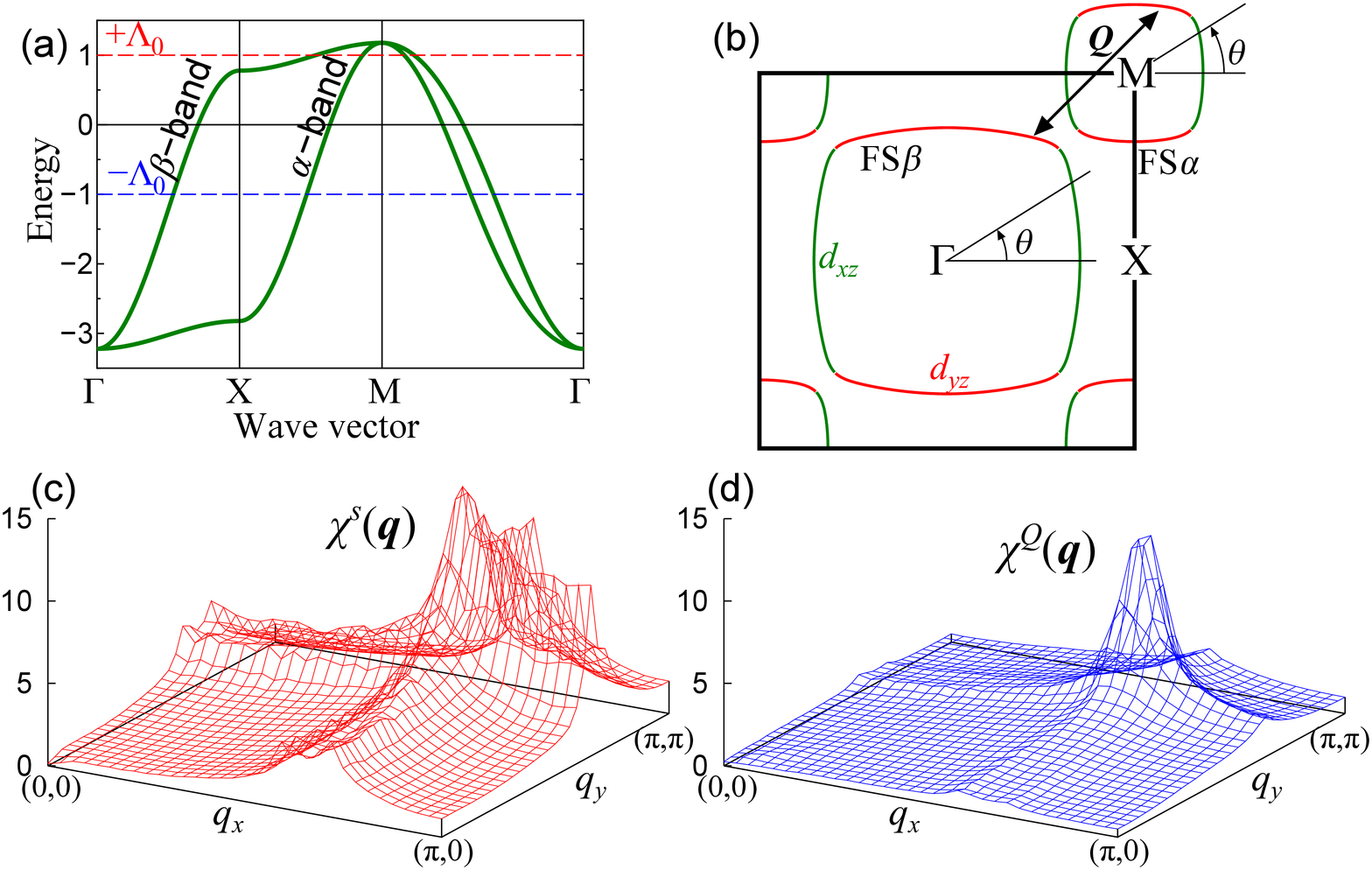}
\caption{(color online)
(a) Band structure and (b) FSs of the two-orbital model.
${\bm Q}\approx (2\pi/3,2\pi/3)$ is the nesting vector. 
(c) $\chi^s({\bm q})$ and (d) $\chi^Q({\bm q})$ of the q1D-band model
obtained by the RG+cRPA method ($\Lambda_0=1$) 
for $U=3.5$, $J/U=0.035$ and $T=0.02$.
}
\label{fig:FS}
\end{figure}

Figures \ref{fig:FS} (a) and (b) 
show the ($d_{xz}$, $d_{yz}$)-orbital bands
and the corresponding quasi-one-dimensional (q1D) Fermi surfaces (FSs),
FS$\alpha$ and FS$\beta$.
In Sr$_2$RuO$_4$, the nesting of the q1D FSs gives 
the experimentally observed AFM fluctuations at ${\bm Q}\approx(2\pi/3,2\pi/3)$.
In addition, there is a $d_{xy}$-orbital band 
that compose the two-dimensional (2D) FS, FS$\gamma$.
(see Figs.\ \ref{fig:AP-FS} (a) and (b).)
If the spin-orbit interaction (SOI) is neglected,
the ($\alpha,\beta$)-bands and $\gamma$-band are 
coupled only via the Coulomb interaction.
Therefore, the superconductivity is expected to be realized mainly in either 
the q1D bands ($|\Delta_{\alpha,\beta}|\gg |\Delta_\gamma|$) or 
the 2D band ($|\Delta_{\alpha,\beta}|\ll |\Delta_\gamma|$).

The mechanisms of the TSC originating mainly from the 2D band
had been proposed in Refs.\ \cite{Nomura,Wang,Arita,Miyake}:
Nomura and Yamada explained the TSC state by using the
higher-order perturbation theory \cite{Nomura},
which is the natural development of the 
Kohn-Luttinger mechanism \cite{KL}.
Recently, a three-orbital Hubbard model had been studied
using a 2D renormalization group (RG) method \cite{Wang},
and it was claimed that 
the $p$-wave gap is realized on the FS$\gamma$ accompanied by the  
development of spin fluctuations at ${\bm q}=(0.19\pi,0.19\pi)$.
Also, charge-fluctuation-mediated TSC was discussed 
by introducing the inter-site Coulomb interaction \cite{Arita}.

On the other hand, it is natural to expect that the 
TSC is closely related to the strong AFM fluctuations 
in the q1D FSs at ${\bm q}\sim{\bm Q}$.
The TSC originating from the q1D FSs had been discussed
by applying the perturbation theory \cite{Kivelson,SOPT}
and random-phase-approximation (RPA) \cite{Takimoto,Ogata}.
Takimoto discussed the orbital-fluctuation-mediated TSC
using the RPA under the condition $U'>U$,
where $U$ ($U'$) is the intra-orbital (inter-orbital) Coulomb interaction
\cite{Takimoto}.
However, in the RPA, the SSC is obtained
under the realistic condition $U\ge U'$ due to strong AFM fluctuations.
When the spin fluctuation is Ising-like,
the TSC may be favored since the pairing 
interaction for the SSC is reduced \cite{Ogata}.
The charge-fluctuation-mediated TSC was also discussed \cite{Kohmoto}.
However, these theories did not clearly explain
why the TSC overwhelms the SSC in Sr$_2$RuO$_4$
despite the presence of strong AFM fluctuations.

To find out the origin of the TSC in Sr$_2$RuO$_4$,
many experimental efforts have been devoted to determine the gap structure,
such as the tunnel junction \cite{Yada},
ARPES, and quasiparticle interference measurements.
Recently, large superconducting gap with $2|\Delta|\approx 5T_{\rm c}$
was observed by the scanning tunneling microscopy measurements
\cite{Firmo}.
The observed large gap would be that on the q1D FSs, 
since the tunneling will be dominated by the  
$(d_{xz},d_{yz})$-orbitals that stand along the $z$-axis,
as clarified in the double-layer compound Sr$_3$Ru$_2$O$_7$ 
\cite{Lee-327}.
Therefore, it is an important challenge
to establish the theory of the TSC based on the q1D-band Hubbard model,
by applying an advanced theoretical method.

In this paper,
we study the multiorbital Hubbard models for Sr$_2$RuO$_4$ 
with realistic parameters $(U>U')$.
To analyze the many-body effect beyond the mean-field-level approximations,
we apply both the RG+cRPA method developed in Ref.\ \cite{Tsuchiizu}
in Sec.\ \ref{sec:RG}, 
and the SC-VC$_\Sigma$ method in Refs.\ \cite{Onari-SCVC,Onari-Hdoped}
in Sec.\ \ref{sec:SCVC}.
Thanks to the vertex correction (VC) for the susceptibility,
which is dropped in the RPA,
strong orbital and spin fluctuations at ${\bm Q}\approx(2\pi/3,2\pi/3)$
emerge in the q1D bands, and it is confirmed that 
the TSC is efficiently induced by the cooperation of both fluctuations
for wide parameter region.
The idea of the  ``spin+orbital fluctuation mediated TSC state'' 
will be realized not only in Sr$_2$RuO$_4$, 
but also in other triplet superconductors.
In Sec.\ \ref{sec:with-SOI},
we discuss the effect of the SOI on the superconductivity, and 
reveal that the large gap on the FS$\gamma$ is induced 
from the FS$\beta$ due to the SOI-induced ``orbital-mixture'' 
between FS$\beta$ and FS$\gamma$.
This effect will be important to understand the experimental 
approximate $T$-linear behavior of $C/T$ below $T_{\rm c}$ in Sr$_2$RuO$_4$.

\vspace*{-.2cm}

\section{Renormalization Group + cRPA Method}
\label{sec:RG}

\vspace*{-.2cm}

In this section, we analyze this model by applying the RG combined with the 
constrained RPA (RG+cRPA) \cite{Tsuchiizu}.
This method is very powerful to calculate the 
higher-order many-body effects systematically and in an unbiased way.
In the RG+cRPA method, we divide the lower-energy region  ($|E|<\Lambda_0$) 
of the Brillouin zone into $N$ patches as done in 
Refs.\ \cite{Metzner,Honerkamp,RG-Rev,DHLee} and perform the RG analysis. 
The contributions from the higher-energy region ($|E|>\Lambda_0$)
are calculated by the cRPA method with high numerical accuracy, 
and incorporated into the initial vertex functions \cite{Tsuchiizu}.
(The conventional patch-RG method \cite{Metzner,Honerkamp,RG-Rev,DHLee}
is recovered when $\Lambda_0>W_{\rm band}$.)
Although the initial vertex functions are very small,
they play decisive roles for the fixed point of the RG flow.

\vspace*{-.2cm}

\subsection{Analysis of the Two-Orbital Model}
\label{sec:2orbital}

\vspace*{-.2cm}

First, we study the two-orbital Hubbard model,
which describes the quasi-1D FSs of Sr$_2$RuO$_4$.
The kinetic term is given by
\begin{eqnarray}
H_0=\sum_{{\bm k},\sigma}
\sum_{l,m}^{1,2}\xi_{\bm k}^{l,m}c_{{\bm k},l,\sigma}^\dagger c_{{\bm k},m,\sigma},
\end{eqnarray}
where the orbital indices $l,m=1$ and $2$ refer to 
$d_{xz}$- and $d_{yz}$-orbitals, respectively.
In the present model,
$\xi_{\bm k}^{1,1}=-2t\cos k_x-2t_{\rm nn}\cos k_y$,
$\xi_{\bm k}^{2,2}=-2t\cos k_y-2t_{\rm nn}\cos k_x$, and
$\xi_{\bm k}^{1,2}=4t'\sin k_x \sin k_y$.
Hereafter, we set $(t,t_{\rm nn}, t')=(1,0.1,0.1)$, and 
fix the filling as $n=4\cdot (2/3)=2.67$,
which corresponds to the filling of the q1D FSs of Sr$_2$RuO$_4$.
We also introduce the on-site Coulomb interactions $U$, $U'$, 
and put the exchange and Hund's couplings $J=J'=(U-U')/2$ 
throughout the paper.

Here, we analyze this model by applying the RG+cRPA method \cite{Tsuchiizu}.
We use $N=64$ (32 patches for each FS) in the present study, 
and it is verified that the results of $N=128$ are almost unchanged.
First, we calculate the susceptibilities using the RG+cRPA:
The charge (spin) susceptibility is given by
\begin{eqnarray}
&& \!\!\!\!\!\!
\chi_{l,l';m,m'}^{c(s)}(q)= 
\int_0^\beta d\tau
\frac12 \langle A_{l,l'}^{c(s)}({\bm q},\tau)A_{m',m}^{c(s)}(-{\bm q},0)\rangle
e^{i\omega_l\tau},
\quad
 \\
&& {} \,\,\,
A_{l,l'}^{c(s)}({\bm q})=\sum_{{\bm k}} (c^\dagger_{{\bm k},l',\uparrow}c_{{\bm k}+{\bm q},l,\uparrow}
+(-)c^\dagger_{{\bm k},l',\downarrow}c_{{\bm k}+{\bm q},l,\downarrow}) ,
\end{eqnarray}
where 
$q=({\bm q},\omega_l)$, and $l,l',m,m'$ are $d$ orbitals.
The quadrupole susceptibility with respect to
$O_{x^2-y^2}=n_{xz}-n_{yz}$ and the total spin susceptibility
are respectively given as
\begin{eqnarray}
\chi^Q({\bm q})&=& \sum_{l,m}(-1)^{l+m}\chi_{l,l;m,m}^{c}({\bm q}),
\\
\chi^s({\bm q})&=& \sum_{l,m}\chi_{l,l;m,m}^{s}({\bm q}).
\end{eqnarray}
Figures \ref{fig:FS} (c) and (d) show the obtained 
$\chi^s({\bm q})$ and $\chi^Q({\bm q})$, respectively,
by the RG+cRPA method ($\Lambda_0=1$)
for $U=3.5$ and $J/U=0.035$ at $T=0.02$.
Both susceptibilities have the peak at ${\bm Q}\approx(2\pi/3,2\pi/3)$,
which is the nesting vector of the present FSs.
The shape of $\chi^s({\bm q})$ is essentially equivalent to 
that of the RPA, by putting  $U=2.2$ and $J/U=0.035$.
However,  $\chi^Q({\bm q})$ in the RPA is quite small when $J>0$
\cite{Onari-SCVC,Ohno-SCVC}.
Therefore, the enhancement of $\chi^Q({\bm q})$ in Fig.\ \ref{fig:FS} (d)
originates from the many-body effect beyond the RPA.
The natural candidate is the Aslamazov-Larkin (AL) type
VC for $\chi^Q(q)$, $X^c(q)$,
whose analytic expression is given in Ref.\ \cite{Onari-SCVC}.
Since $X^c(q) \sim U^4 T\sum_k \Lambda_{\rm AL}(q;k)^2 \chi^s(k)\chi^s(k+q)$
for simplicity, $X^c(q)$ takes large value at ${\bm q}={\bm 0}$ and $2{\bm Q}$
when $\chi^s({\bm k})$ is large at ${\bm k}={\bm Q}$.
$\Lambda_{\rm AL}(q;k)$ is the three-point vertex
composed of three Green functions \cite{Onari-SCVC}.
In the present model, $2{\bm Q}\approx{\bm Q}$ in the first Brillouin zone.
Thus, with the aid of the VC and the nesting of the FSs,
the enhancement of $\chi^Q({\bm Q})$ in Fig.\ \ref{fig:FS} (d) is realized.

\begin{figure}[t]
\includegraphics[width=.9\linewidth]{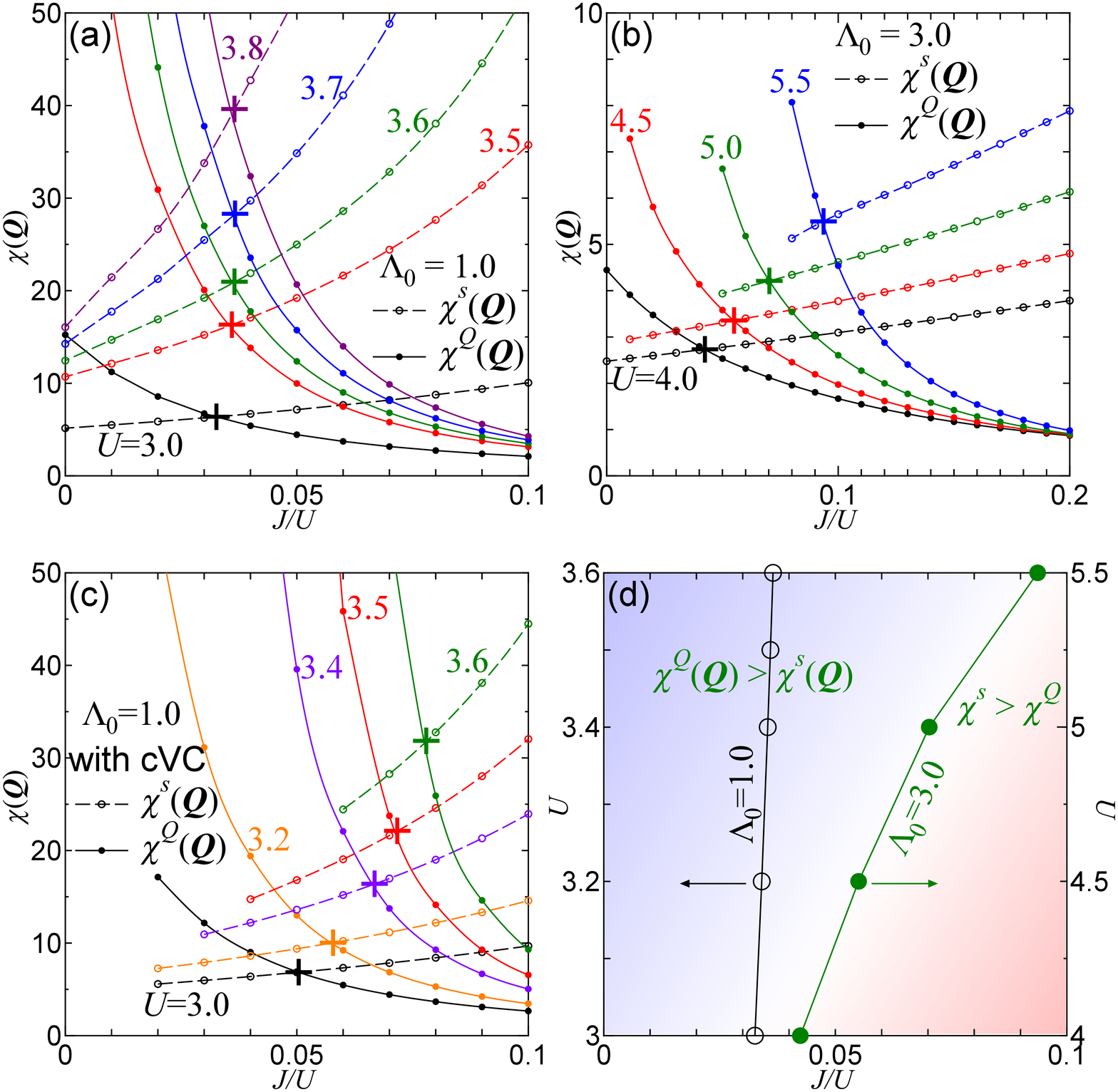}
\caption{(color online)
$\chi^s({\bm Q})$ and $\chi^Q({\bm Q})$ as functions of $J/U$
given by the RG+cRPA method for (a) $\Lambda_0=1$ ($U=3.0\sim3.8$)
and (b) $\Lambda_0=3$ ($U=4.0\sim5.5$).
(c) $\chi^s({\bm Q})$ and $\chi^Q({\bm Q})$ for $\Lambda_0=1$,
by including the cVC.
(d) Obtained phase diagram for $\Lambda_0=1$
and $\Lambda_0=3$ without cVC.
}
\label{fig:chiQ}
\end{figure}

Figure \ref{fig:chiQ} (a) shows $\chi^s({\bm Q})$ and $\chi^Q({\bm Q})$ 
as functions of $J/U$ at $T=0.02$, 
obtained by the RG+cRPA method with $\Lambda_0=1$.
For each value of $U$,
$\chi^Q({\bm Q})$ ($\chi^s({\bm Q})$) decreases (increases) with $J/U$,
and they are equal at $(J/U)_{c}\sim0.035$.
We stress that $(J/U)_{c}$ is negative in the RPA
since the VC is totally dropped.
In the case of $\Lambda_0=3$ shown in Fig.\ \ref{fig:chiQ} (b), 
the value of $(J/U)_{c}$ 
increases to $\sim0.09$ at $U\sim5.5$, indicating that importance 
of the VC due to higher energy region.
To check this expectation, we include the 
constrained AL term (cVC) in addition to the cRPA \cite{Tsuchiizu}.
The obtained results are shown in Fig.\ \ref{fig:chiQ} (c).
It is verified that $(J/U)_{c}$ increases to $0.08$ at $U=3.6$.
($\chi^Q({\bm Q})$ in Fig.\ \ref{fig:chiQ} (c) is approximately given by 
shifting $\chi^Q({\bm Q})$ in Fig.\ \ref{fig:chiQ} (a) 
horizontally by $+0.02\sim+0.05$.)
The values of $(J/U)_{c}$ obtained by Figs.\ \ref{fig:chiQ} (a)-(c) 
are summarized in Fig.\ \ref{fig:chiQ} (d).
We find that $(J/U)_{c}\sim0.15$ in the self-consistent VC method
with the self-energy correction (SC-VC$_{\Sigma}$ method)
\cite{Onari-Hdoped}.

\begin{figure}[!htb]
\includegraphics[width=.9\linewidth]{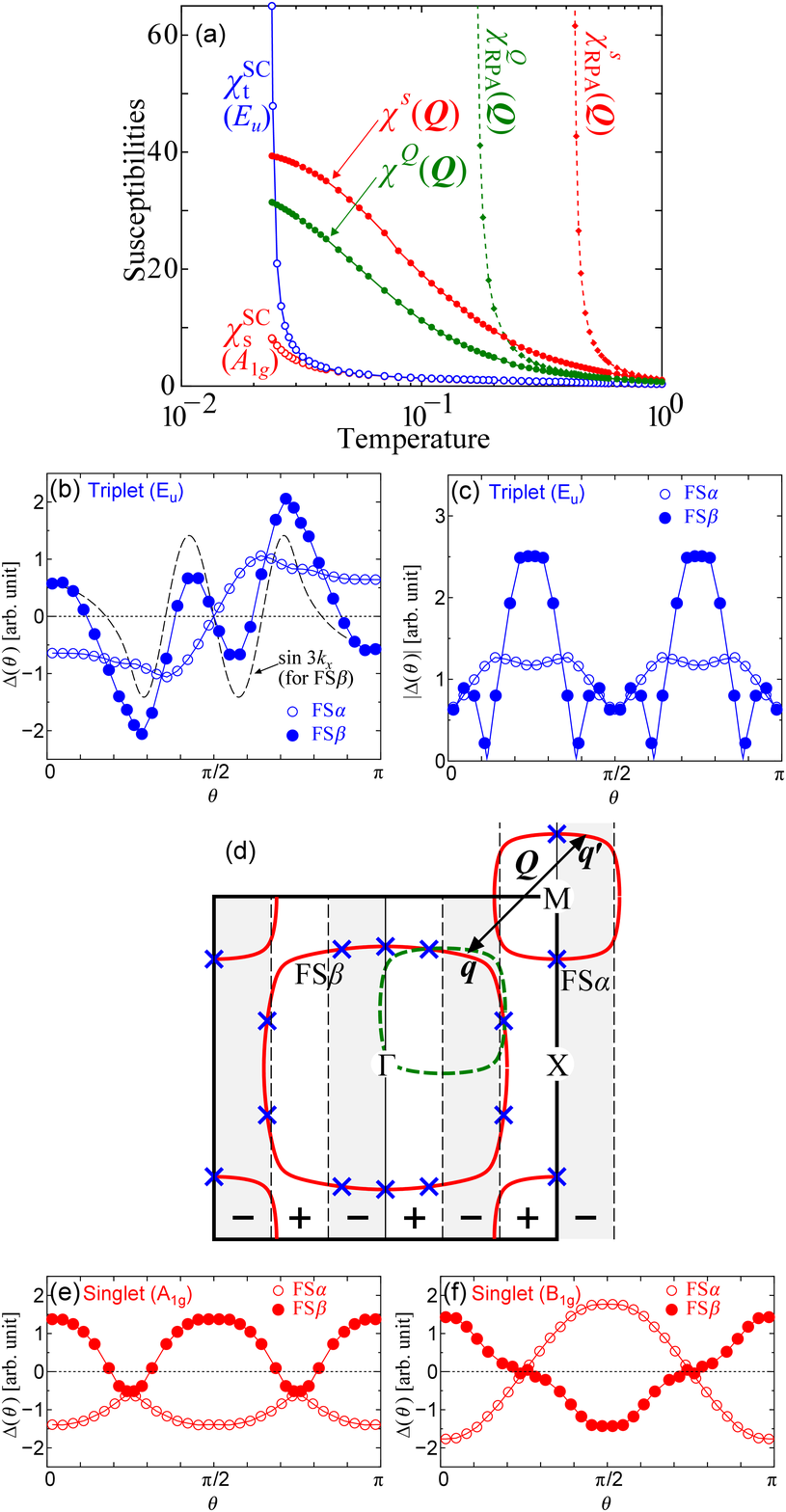}
\caption{(color online)
(a) $T$-dependences of $\chi^s({\bm Q})$, $\chi^Q({\bm Q})$,
$\chi_{\rm s}^{\rm SC}$ and $\chi_{\rm t}^{\rm SC}$ for $U=3.8$ and $J/U=0.04$
($\Lambda_0=1$).
(b) $E_u$ gap functions on FS$\mu$, $\Delta_x^\mu(\theta)$
($\mu=\alpha,\beta$) obtained by the RG.
The relation $\Delta_x^\beta \propto \sin 3k_x$ holds approximately.
$N=128$ patches are used.
(c) The magnitude of the chiral (or helical) gap state
$|\Delta^\mu|= \sqrt{(\Delta_x^\mu)^2+(\Delta_y^\mu)^2}$.
(d) Schematic explanation for the $\sin 3k_x$-type TSC due to 
orbital+spin fluctuations at ${\bm q}={\bm Q}$.
Solid lines (broken lines) are the necessary (accidental) nodes.
The positions of nodes ($\Delta_x^\mu=0$) in (b) are shown by crosses.
(e) $A_{1g}$ and (f) $B_{1g}$ SSC gap functions.
}
\label{fig:SC}
\end{figure}

Although the value of $(J/U)_{c}$ is underestimated at $\Lambda_0=1$,
the obtained $\chi^s({\bm q})$ and $\chi^Q({\bm q})$ at $\Lambda_0=1$ are reliable,
since the higher-energy processes can be calculated
with high numerical accuracy \cite{Tsuchiizu}.
Hereafter, we perform the RG+cRPA method with $\Lambda_0=1$,
by using smaller $J/U\ (\sim0.04)$ to 
compensate for the absence of the higher-energy VCs.
Figure \ref{fig:SC} (a) shows the $T$-dependences of
$\chi^s({\bm Q})$ and $\chi^Q({\bm Q})$ given by the RG+cRPA method
($\Lambda_0=1$) for $U=3.8$ and $J/U=0.04$:
Both of them are strongly renormalized from the RPA results.
In the RPA, $\chi^s_{\rm RPA}({\bm Q})$ diverges at $T\approx 0.4$,
at which $\chi^Q_{\rm RPA}({\bm Q})$ remains very small.
In highly contrast, in the RG+cRPA method,
the relation $\chi^s({\bm Q}) \approx \chi^Q({\bm Q})$ holds for wide temperature range.

We also calculate the TSC and SSC susceptibilities
using the RG+cRPA method:
\begin{eqnarray}
\chi_{\rm t(s)}^{\rm SC}=
\frac12 \int_0^\beta d\tau\langle B_{\rm t(s)}^\dagger(\tau)B_{\rm t(s)}(0)\rangle,
\label{eqn:chiSC}
\end{eqnarray}
where $B_{\rm t(s)} = 
\sum_{{\bm q},\mu}\Delta_{\rm t(s)}^\mu({\bm q}) c_{{\bm q},\mu,\uparrow} c_{-{\bm q},\mu,\uparrow(\downarrow)}$.
$\mu=\alpha,\beta$ is the band index, and
$\Delta_{\rm t(s)}^\mu({\bm q})$ is the odd (even) parity gap function.
The obtained $\chi_{\rm t(s)}^{\rm SC}$ is shown in Fig.\ \ref{fig:SC} (a),
by optimizing the functional form of $\Delta_{\rm t(s)}^\mu({\bm q})$ numerically
\cite{comment-d}.
Since $\chi_{\rm t(s)}^{\rm SC}$ diverges at $T=T_{\rm c}$, 
the strong development of $\chi_{\rm t}^{\rm SC}$ at $T\approx 0.02$
means that the TSC is realized.
This TSC state belongs to the two-dimensional $E_u$-representation,
$(\Delta_{x}^\mu({\bm q}), \Delta_{y}^\mu({\bm q}))$.
The obtained $\Delta_{x}^\mu$ on the FSs when $\chi^{\rm SC}_{\rm t}\sim60$
are shown in Fig.\ \ref{fig:SC} (b), where $\theta$ is the 
angle of the Fermi momentum shown in Fig.\ \ref{fig:FS} (b).
The necessary nodes $\Delta_{x(y)}^\mu=0$ 
are on the lines $q_{x(y)}=0,\pm\pi$.
Very similar TSC gap is obtained for $J/U\lesssim0.08$
by taking the cVC into account with $\Lambda_0=1$.
Below $T_{\rm c}$, the BCS theory tells that the
chiral or helical gap state with the gap amplitude
$|\Delta^\mu|= \sqrt{(\Delta_x^\mu)^2+(\Delta_y^\mu)^2}$,
which is shown in Fig.\ \ref{fig:SC} (c),
is realized to gain the condensation energy.

To understand why the TSC state is obtained,
it is useful to analyze the linearized gap equation:
\begin{eqnarray}
\lambda_{a}^E {\bar \Delta}_{a}^\mu({\bm q})
&=& -\sum_{\mu'}^{\alpha,\beta} \int_{\rm FS\mu'}\frac{d{\bm q}'}{v_{{\bm q}'}^{\mu'}}
V_{a}^{\mu,\mu'}({\bm q},{\bm q}') {\bar \Delta}_{a}^{\mu'}({\bm q}')
 \nonumber\\
& &{} \times \ln (1.13\omega_c/T),
\label{eqn:GapEq}
\end{eqnarray}
where $a = t$ or $s$.
$\lambda^E_{a}$ is the eigenvalue, 
$V_{a}^{\mu,\mu'}({\bm q},{\bm q}')$ is the pairing interaction, and
$\omega_c$ is the cut-off energy of the interaction.
As shown in Fig.\ \ref{fig:FS} (b), 
the inter-band interaction ($\mu=\alpha$, $\mu'=\beta$)
with ${\bm q}-{\bm q}'={\bm Q}$ is approximately given by the intra-orbital interaction
given as
\begin{eqnarray}
{V}_{a}^l({\bm q},{\bm q}')
&=& b_{a}\frac{U^2}{2} |{\Lambda}_l^s({\bm q};{\bm q}')|^2
{\chi}_l^s({\bm q}-{\bm q}') 
\nonumber \\
& &{} + c_{a}\frac{U^2}{2} |{\Lambda}_l^c({\bm q};{\bm q}')|^2
{\chi}_l^c({\bm q}-{\bm q}') ,
 \label{eqn:Vts} 
\end{eqnarray}
where $(b_{\rm t},c_{\rm t})=(-1,-1)$ and $(b_{\rm s},c_{\rm s})=(3,-1)$,
and ${\chi}_l^{s,c}({\bm Q})\equiv {\chi}_{l,l;l,l}^{s,c}({\bm Q})$.
(Note that ${\chi}_l^s({\bm Q})\approx{\chi}^s({\bm Q})/2$ 
and ${\chi}_l^c({\bm Q})\approx{\chi}^Q({\bm Q})/4$,
since ${\chi}_{l}^{s}({\bm Q})\gg {\chi}_{1,1;2,2}^{s}({\bm Q})$
and ${\chi}_{l}^{c}({\bm Q})\approx -{\chi}_{1,1;2,2}^{c}({\bm Q})$
near the critical point \cite{Ohno-SCVC}.)
${\Lambda}_l^{s,c}$ is the VC for the gap equation,
which we call $\Delta$-VC in Ref.\ \cite{Onari-Hdoped}.
The AL-type diagram for the charge channel is given by
${\Lambda}_l^{c}({\bm q};{\bm q}') \sim
1+T\sum_k \Lambda_{\rm AL}(q-q';k)G(k) \chi^s(k+q)\chi^s(k-q')$,
which is strongly enlarged for ${\bm q}-{\bm q}'\approx {\bm Q}$,
and the orbital-fluctuation-mediated pairing is favored
\cite{Onari-SCVC,Onari-Hdoped}. 
The merit of the RG+cRPA method is that the
the  AL-type $\Delta$-VC is automatically produced
in calculating the pairing susceptibility in Eq. (\ref{eqn:chiSC}).

In the RPA with $J>0$,
the TSC cannot be achieved because of the relation 
$\chi_l^s({\bm Q}) \gg \chi_l^c({\bm Q})$ and $\Lambda^{c,s}=1$ in the RPA:
In this case, spin-fluctuation-mediated SSC is obtained since
$|{V}_{\rm s}^l|\approx 
 (U^2/2)\{3|\Lambda_l^s|^2\chi_l^s-|\Lambda_l^c|^2\chi_l^c\}$ 
is three times larger than 
$|{V}_{\rm t}^l|\approx
 (U^2/2)\{|\Lambda_l^s|^2\chi_l^s+|\Lambda_l^c|^2\chi_l^c\}$.
In the present RG+cRPA method, in contrast, the relationship
$\chi_l^s({\bm Q}) \sim \chi_l^c({\bm Q})$ is realized, and therefore
the triplet interaction $|{V}_{\rm t}^l|$
can be larger than $|{V}_{\rm s}^l|$.
We verified that the TSC susceptibility overcomes the 
$s$-wave and $d$-wave SSC susceptibilities
when $\chi^Q({\bm Q})$ is comparable or larger than $\chi^s({\bm Q})$, 
which is realized in wide parameter range in the present RG+cRPA method 
as shown in Figs.\ \ref{fig:chiQ} (a)-(c).

Using Fig.\ \ref{fig:SC} (d), we explain the gap structure of the 
TSC state induced by orbital+spin fluctuations at ${\bm q}\approx{\bm Q}$.
In addition to the necessary nodes shown by solid lines, 
accidental nodal lines appear around
$k_x\approx \pm\pi/3$ and $k_x\approx \pm2\pi/3$:
The reason is that $\Delta_{x}^\alpha({\bm q})$ and $\Delta_{x}^\beta({\bm q}')$ 
tend to have the same sign for ${\bm q}-{\bm q}'\approx {\bm Q}$
due to large attractive interaction by $V_{t}^l({\bm q}, {\bm q}')$.
For this reason, the relation 
$\Delta_{x}^\beta({\bm q}) \sim \sin 3k_x$ in Fig.\ \ref{fig:SC} (b)
is satisfied in the $E_u$-type TSC state.

In Fig.\ \ref{fig:SC} (a),
$\chi_{\rm s}^{\rm SC}$ also develops at low temperatures:
Figures  \ref{fig:SC} (e) and (f) show the obtained 
$A_{1g}$ and $B_{1g}$ SSC gap structures, which give
the first and the second largest $\chi_{\rm s}^{\rm SC}$'s.
Both SSC states with sign reversal are mainly caused by 
spin fluctuations, and $A_{1g}$ state is slightly stabilized 
by the orbital fluctuations.
The $A_{1g}$ state in Fig.\ \ref{fig:SC} (e) dominates the TSC state
when $\chi^s({\bm Q})\gg\chi^Q({\bm Q})$, which is realized for $J/U\gtrsim0.05$
in Fig.\ \ref{fig:chiQ} (a).

\begin{figure}[t]
\includegraphics[width=.9\linewidth]{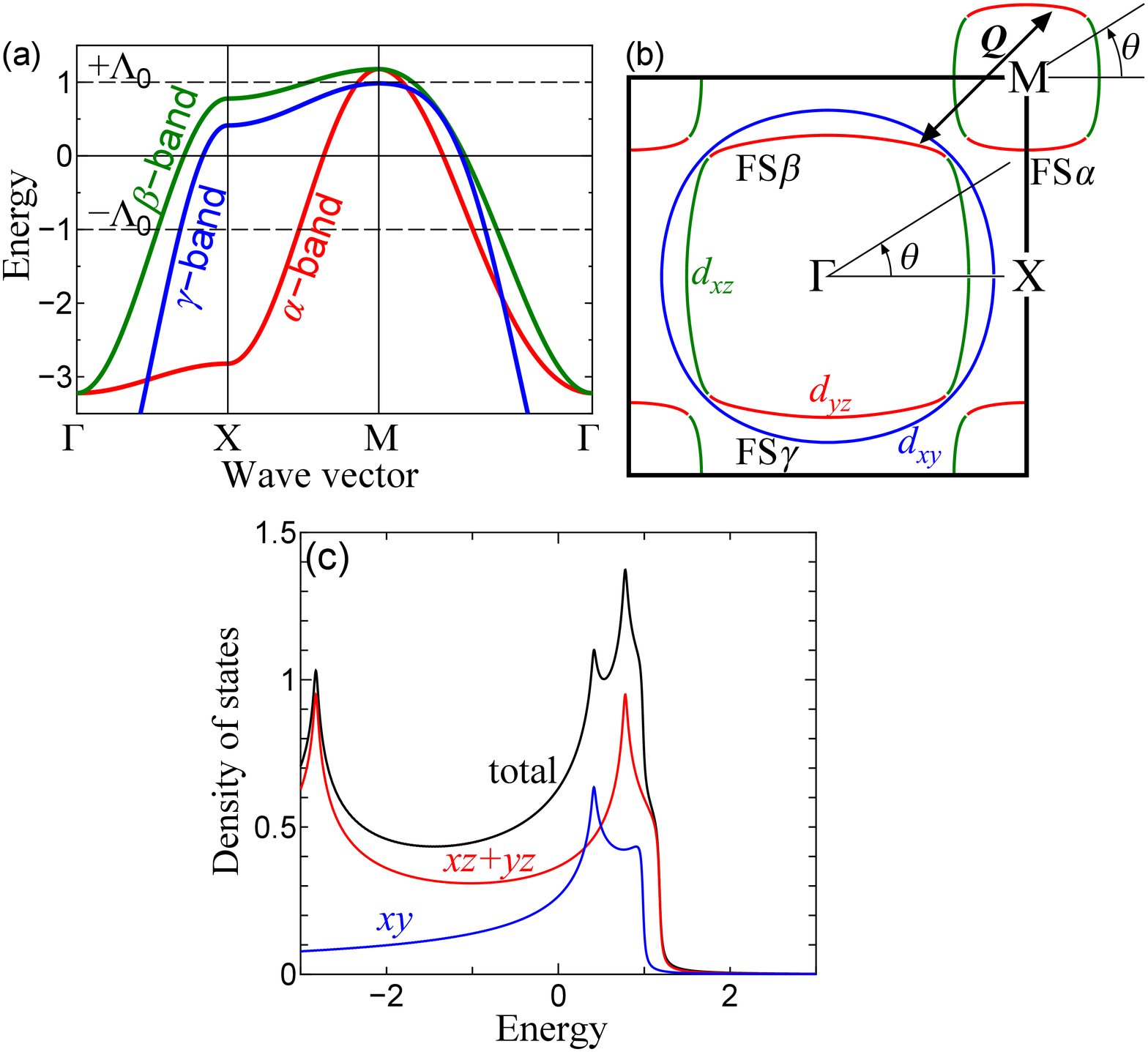}
\caption{(color online)
(a) Band structure, (b) three FSs, and 
(c) DOS in the three-orbital model for Sr$_2$RuO$_4$ without the SOI.
In the present model, 
$N_{xy}(0)/N_{\rm total}(0)=0.42$,
which is consistent with the value of
the band calculation ($0.47$).
}
\label{fig:AP-FS}
\end{figure}

\begin{figure}[b]
\includegraphics[width=.9\linewidth]{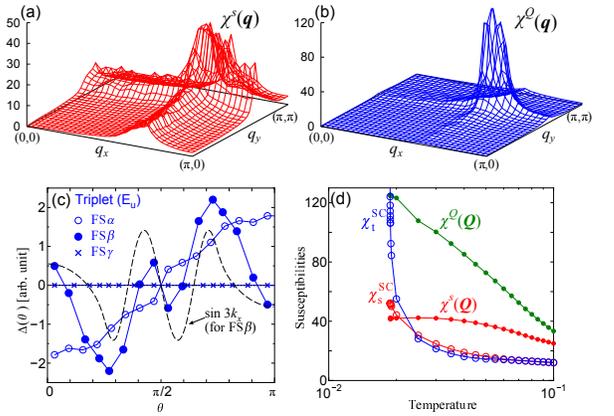}
\caption{(color online)
(a) $\chi^s({\bm q})$ and (b) $\chi^Q({\bm q})$ 
given by applying the RG+cRPA method to the three-orbital model.
(c) The TSC gap functions as function of $\theta$.
(d) Obtained temperature dependence of the TSC (SSC) susceptibility,
$\chi_{\rm t(s)}^{\rm SC}$, together with $\chi^Q({\bm Q})$ and $\chi^s({\bm Q})$.
}
\label{fig:AP-RG}
\end{figure}

\vspace*{-.2cm}

\subsection{Analysis of the Three-Orbital Model}
\label{sec:3orbital}

\vspace*{-.2cm}

In the previous subsection, we showed that the
``spin+orbital fluctuation mediated TSC state''
emerges by analyzing the two-orbital Hubbard model.
In order to verify this result,
we study a more realistic three orbital Hubbard model for Sr$_2$RuO$_4$:
We add the $\gamma$-band given by the $d_{xy}$-orbital,
$\xi_{\bm k}^3= -2t_3 (\cos k_x + \cos k_y) -4t_3'\cos k_x \cos k_y + E_3$,
to the present two-orbital model \cite{Nomura,Wang}.
We put $t_3=0.86$, $t_3'=0.36$, $E_3=0.01$, $n_{xz}+n_{yz}=2.67$ and 
$n_{xy}=1.28$.
In Fig.\ \ref{fig:AP-FS}, we show the (a) band structure,
(b) FSs, and (c) density-of-states (DOS) 
in the present three-orbital model in the absence of the SOI.

\begin{figure}[b]
\includegraphics[width=.9\linewidth]{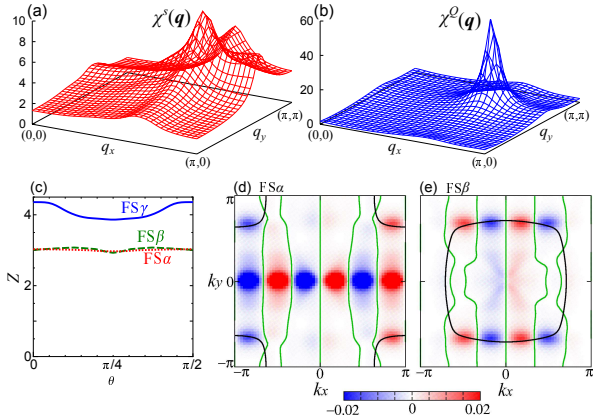}
\caption{(color online)
(a)  $\chi^s({\bm q})$, (b) $\chi^Q({\bm q})$, and
(c) $Z_{\delta}$ ($\delta=\alpha,\beta,\gamma$) given by the SC-VC$_\Sigma$ method
for the three-orbital model without SOI.
The obtained relation $Z_{\alpha,\beta}/Z_{\gamma}=0.7$ is consistent
with the result given by the FLEX approximation \cite{Arakawa}.
(d),(e) The TSC gap functions on $\alpha$-band and $\beta$-band
obtained by the SC-VC$_\Sigma$ method.
}
\label{fig:AP-SCVCS}
\end{figure}

We analyze this three-orbital Hubbard model
by applying the RG+cRPA method.
Here, we take the renormalization of the Green function due to $1/Z_\delta$
($\delta=\alpha,\beta,\gamma$) according to Ref.\ \cite{renormalization},
where $Z_\delta$ is the mass-enhancement constant of $\delta$-band, 
and we put $Z_\gamma/Z_{\alpha,\beta}=1.4$ ($Z_{\alpha,\beta}=1$)
as observed by ARPES and dHvA measurements \cite{Borisenko},
and as theoretically obtained by the SC-VC$_\Sigma$ method
(see Fig.\ \ref{fig:AP-SCVCS} (c)).
Figures \ref{fig:AP-RG} (a) and (b) show the  
obtained $\chi^s({\bm q})$ and $\chi^Q({\bm q})$, respectively.
We put $\Lambda_0=1$, $U=3.9$ and $J/U=0.02$.
Thus, the strong orbital and spin fluctuations at 
${\bm Q}\approx(2\pi/3,2\pi/3)$ due to q1D bands are obtained
in the three-orbital model.
The wavenumber of the incommensurate peak of $\chi^s({\bm q})$
is consistent with the neutron measurements.

We also study the superconducting susceptibility, and
the obtained TSC gap functions on FS$\alpha$ and FS$\beta$ are
shown in Fig.\ \ref{fig:AP-RG} (c).
These obtained results are very similar to those in the two-orbital model.
The TSC gap function on FS$\gamma$ is very small
since the orbital fluctuations do not develop in $d_{xy}$-orbital.
The temperature dependences of the TSC (SSC) susceptibilities,
$\chi_{\rm t(s)}^{\rm SC}$, in  Fig.\ \ref{fig:AP-RG} (d),
together with $\chi^Q({\bm Q})$ and $\chi^s({\bm Q})$.
Therefore, the spin and quadrupole susceptibilities
as well as the TSC gap function obtained in the three-orbital Hubbard model  
are very similar to those given in the two-orbital model.

\vspace*{-.2cm}

\section{Self-Consistent Vertex-Correction (SC-VC) method:
Analysis of the Three-Orbital Model
}
\label{sec:SCVC}

\vspace*{-.2cm}

In this section, we analyze the three-orbital Hubbard model by using 
the SC-VC method with the self-energy correction (SC-VC$_\Sigma$ method) 
\cite{Onari-SCVC,Onari-Hdoped}
in order to confirm the validity of the results of the RG+cRPA method 
shown in Sec.\ \ref{sec:3orbital}.
In this section, we set the charge Stoner factor $\sim0.98$
in the SC-VC$_\Sigma$ method.
In Fig.\ \ref{fig:AP-SCVCS}, we show the obtained 
(a) spin susceptibility $\chi^s({\bm q})$, 
(b) quadrupole susceptibility $\chi^Q({\bm q})$,
and (c) mass-enhancement factor for FS-$\delta$ ($\delta=\alpha,\beta,\gamma$),
$Z_\delta({\bm k})= 1-\left.{\rm Re}
\frac{\partial}{\partial\epsilon}\Sigma_\delta({\bm k},\epsilon)\right|_{\epsilon=0}$,
in the case of $U=3.05$ and $J/U=0.12$ at $T=0.02$.
We obtain the relation $Z_{\alpha,\beta}/Z_\gamma\sim0.7$, which is consistent with the
dHvA and ARPES measurements \cite{Borisenko}
as well as the result of the FLEX approximation \cite{Arakawa}.
(The mass-enhancement is mainly caused by spin-fluctuations,
because of the factor $3$ in 
$\Sigma(k) \sim \sum_q G(k+q) \frac{1}{2}U^2(3\chi^s(q) + \chi^c(q))$
in the SC-VC$_\Sigma$ method.)
Since the spin fluctuations in the $\gamma$-band is suppressed
by the large $Z_\gamma({\bm k})$, we obtain the
experimental strong spin fluctuations at 
${\bm Q}\approx(2\pi/3,2\pi/3)$ due to q1D bands
\cite{Arakawa}.

\vspace*{-.2cm}

\subsection{Gap Equation without the SOI:
Verification of the orbital+spin fluctuation mediated TSC
using the \\
SC-VC$_\Sigma$ method
}
\label{sec:no-SOI}

\vspace*{-.2cm}

Here, we analyze the linearized gap equation.
For the triplet ($a=t$) and singlet ($a=s$) superconductivity,
the gap equation in the absence of the SOI is given as
\begin{eqnarray}
\lambda_{a}^E {\bar \Delta}_{a}^\mu(q)
&=& -T\sum_{q'}\sum_{\mu'}^{\alpha,\beta,\gamma} V_{a}^{\mu,\mu'}(q,q') 
\nonumber \\
& &\times
|G_{\mu'}(q')|^2{\bar \Delta}_{a}^{\mu'}(q') ,
\end{eqnarray}
where 
$G_{\mu}(q)$ is the Green function for band $\mu$ with self-energy correction, 
and $q=({\bm q},\epsilon_n)$, where $\epsilon_n$ is the fermion Matsubara frequency.
Also, 
\begin{eqnarray}
V_{a}^{\mu,\mu'}(q,q')&=& 
\sum_{l,l',m,m'} 
U_{\mu;l}({\bm q}) \, 
U_{\mu;l'}(-{\bm q}) \,
U_{\mu';m}({\bm q}')^*
\nonumber \\
&& {} \times
U_{\mu';m'}(-{\bm q}')^* \,
V_{a}^{l,m;m',l'}(q-q'),
\end{eqnarray}
where $U_{\mu;l}({\bm k})=\langle \mu,{\bm k}| l,{\bm k} \rangle$ is the unitary matrix.
$l,l',m,m'$ represent the $d$-orbital.
${\hat V}_{a}(q)= b_{a}{\hat V}^s(q)+ c_{a}{\hat V}^c(q)$ 
is the pairing interaction 
with the vertex correction for the SC gap ($\Delta$-VC)
in the orbital basis, given in Ref.\ \cite{Onari-Hdoped}.
For $U=3.05$ and $J/U=0.12$ at $T=0.05$,
the eigenvalue of the TSC state, $\lambda_{\rm t}^E=0.304$,
is larger than that of the SSC state,
$\lambda_{\rm s}^E=0.291$.
Figures \ref{fig:AP-SCVCS} (d) and (e)
show the obtained TSC gap functions on $\alpha$-band and $\beta$-band
at the lowest Matsubara frequency.
The relation $\Delta_x^{\alpha,\beta} \propto \sin 3k_x$ is satisfied approximately.
Without the SOI, the 2D band does not contribute to the TSC 
since the relation $|\Delta_x^\gamma({\bm k})|/|\Delta_x^{\alpha,\beta}({\bm k})|\ll 0.1$ holds.

We also performed the SC-VC$_\Sigma$ study for $U=3.2$ and $J/U=0.15$,
and verified that the obtained numerical results are similar to
Figs.\ \ref{fig:AP-SCVCS} (a)-(e) for $U=3.05$ and $J/U=0.12$.
Especially, the relationship $\lambda_{\rm t}^E>\lambda_{\rm s}^E$ is satisfied,
and the ($\sin 3k_x$)-like TSC gap function is obtained.

Thus, we obtained the strong orbital+spin fluctuations at 
${\bm Q}\approx(2\pi/3,2\pi/3)$ as well as the TSC state in the q1D bands,
by analyzing the three-orbital model 
using the SC-VC$_\Sigma$ method.
These results are consistent with the results 
given by the RG+cRPA method in Sec.\ \ref{sec:RG}.

\begin{figure}[t]
\includegraphics[width=.9\linewidth]{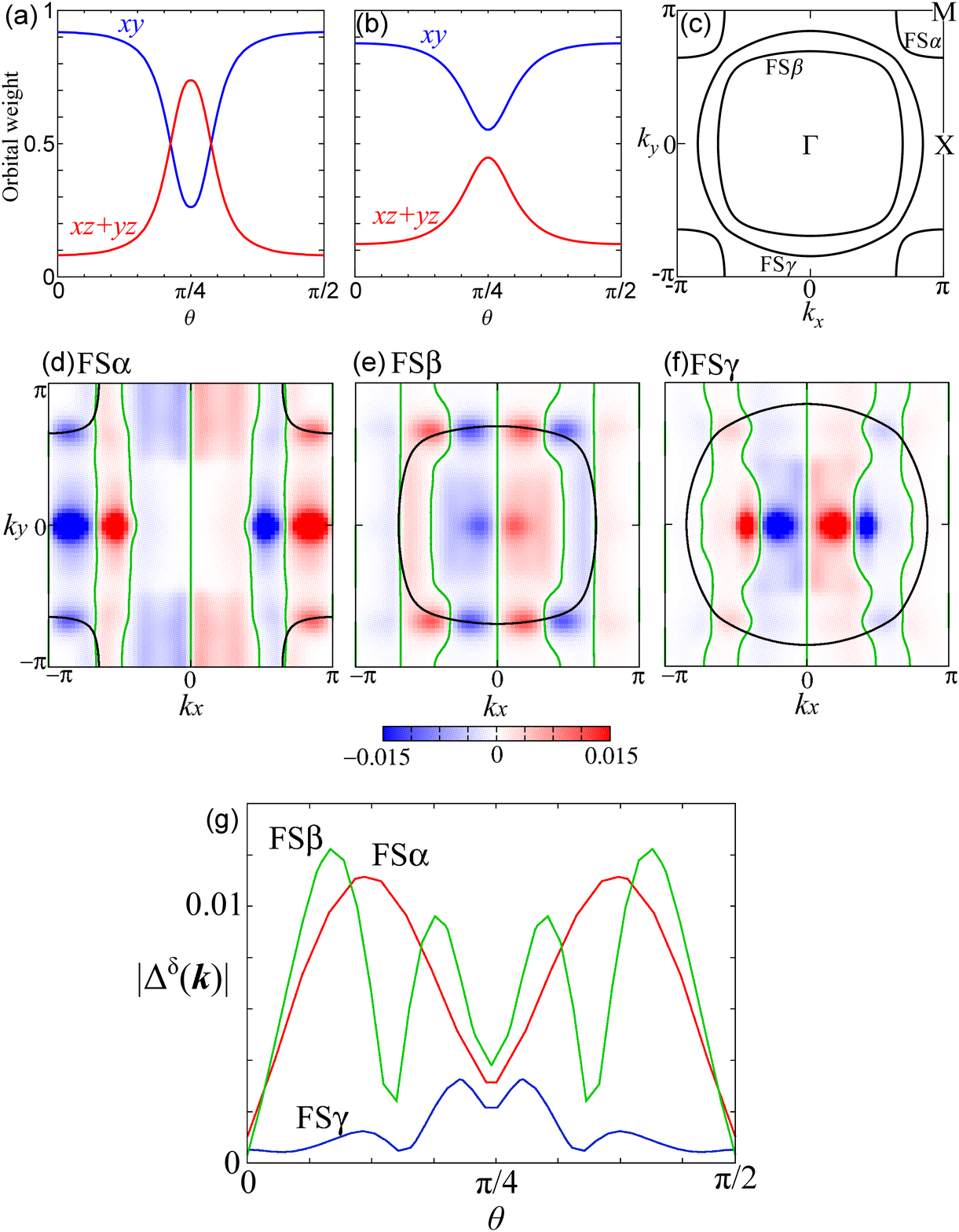}
\caption{(color online)
(a) $d_{xy}$ and $d_{xz+yz}$ orbital weights on the $\gamma$-FS
given by the WIEN2k band calculation with the SOI.
Due to the SOI, $d_{xz+yz}$ orbitals are mixed in the FS$\gamma$
around $\theta=\pi/4$.
(b) Orbital weights on the $\gamma$-FS and (c) three FSs
in the three-orbital model with SOI ($\lambda=0.4$).
(d)-(f) The $E_u$ TSC gap functions $\Delta_x^\delta({\bm k}) \propto \sin 3k_x$
($\delta=\alpha,\beta,\gamma$) obtained by the SC-VC$_\Sigma$ method.
We put $\lambda=0.4$ in the gap equation.
$\Delta_y^\delta(k_x,k_y) \equiv \Delta_x^\delta(k_y,k_x)$ are not shown.
(g) The magnitude of the chiral TSC gap
$|\Delta^\delta({\bm k})|=\sqrt{\Delta_x^\delta({\bm k})^2+\Delta_y^\delta({\bm k})^2}$.
}
\label{fig:AP-SOI}
\end{figure}

\vspace*{-.2cm}

\subsection{Gap Equation with the SOI:
SOI-induced large gap on the FS$\gamma$ due to orbital-mixture
}
\label{sec:with-SOI}

\vspace*{-.2cm}

In this subsection, we discuss the important role of the SOI 
in the superconducting state.
If the SOI is neglected,
the relation $|\Delta_x^\gamma({\bm k})|/|\Delta_x^{\alpha,\beta}({\bm k})|\ll 0.1$ holds
in both the SC-VC$_\Sigma$ and RG+cRPA methods.
However, the experimental relation $C/T\sim T$ below $T_c$ 
($C$ is the specific heat) indicates that 
$\max\{|\Delta_x^\gamma({\bm k})|\}$ takes large value.
We expect that large $\Delta_x^\gamma({\bm k})$ is induced 
from the gap on the FS$\beta$, due to the 
``mixing of the orbital character'' caused by the SOI.
Figure \ref{fig:AP-SOI} (a) shows the orbital weights
$r_{xz+yz}\equiv N_{xz+yz}^\gamma(0)/N_d^\gamma(0)$ 
and $N_{xy}^\gamma(0)/N_d^\gamma(0)$ on the FS$\gamma$
obtained by the WIEN2k with SOI, where $N_l^\gamma(0)$ is the DOS of the 
$l$-orbital ($l=xz, yz, xy$) and $N_d^\gamma(0)=N_{xz+yz}^\gamma(0)+N_{xy}^\gamma(0)$.
Thus, the SOI-induced orbital-mixture is very large,
as reported by the first-principles study \cite{Dama-LDA} and 
observed by Spin-ARPES measurements \cite{Dama-ARPES}.
This result means that the induced gap on the $\gamma$-band,
which is approximately given as
$|\Delta_x^\gamma({\bm k})| \sim r_{xz+yz}({\bm k})|\Delta_x^\beta({\bm k})|$,
becomes comparable to other gaps around $\theta\sim\pi/4$.

Here, we introduce the SOI 
$H_{\rm SOI}= \lambda \sum_i {\bm l}_i\cdot {\bm s}_i$
to the present three-orbital model, and put $\lambda=0.4$.
Figure \ref{fig:AP-SOI} (b) shows the orbital weights
$N_{xy}^\gamma(0)/N_d^\gamma(0)$ and $N_{xz+yz}^\gamma(0)/N_d^\gamma(0)$
on the FS$\gamma$ in the present three-orbital model with $\lambda=0.4$.
The FSs for  $\lambda=0.4$ are shown in Fig.\ \ref{fig:AP-SOI} (c).
By taking the SOI into account, the linearized gap equation 
in terms of the SC-VC$_\Sigma$ method 
for $a=t$, $s$ is given as
\begin{eqnarray}
\lambda^E_{a} {\bar \Delta}_{a,\rho_1 \rho_2}^\mu(q)
&=& -T\sum_{q'}\sum_{\mu'}^{\alpha,\beta,\gamma} 
\sum_{\rho_3,\rho_4}
V_{\rho_1\rho_2;\rho_3\rho_4}^{\mu,\mu'}(q,q') 
\nonumber \\
& &\times
|G_{\mu'}(q')|^2{\bar \Delta}_{a,\rho_3\rho_4}^{\mu'}(q') ,
\end{eqnarray}
where $\mu$, $\mu'$ represent the bands with SOI,
and $\rho_i$ is the pseudo-spin ($\Uparrow,\Downarrow$)
that represents the Kramers doublet.
${\bar \Delta}_{a, \rho_1 \rho_2}^\mu(q)$ is even (odd) 
with respect to the exchange of the pseudospin for $a=t$ ($s$).
Using the ${\bm d}$-vector, the TSC gap functions are expressed as
\begin{eqnarray}
&&\left(
    \begin{array}{cc}
      {\bar \Delta}_{{\rm t}, \Uparrow \Uparrow}^\mu(q) 
& {\bar \Delta}_{{\rm t}, \Uparrow \Downarrow}^\mu(q) \\
      {\bar \Delta}_{{\rm t}, \Downarrow \Uparrow}^\mu(q) 
& {\bar \Delta}_{{\rm t}, \Downarrow \Downarrow}^\mu(q) 
    \end{array}
  \right) \nonumber \\
&&\ \ \ \ \ \ \ \ =
\left(
    \begin{array}{cc}
      -d_x^\mu(q)+id_y^\mu(q) & d_z^\mu(q) \\
      d_z^\mu(q) & d_x^\mu(q)+id_y^\mu(q)
    \end{array}
  \right) .
\end{eqnarray}
The pairing interaction is given as
\begin{eqnarray}
V_{\rho_1\rho_2;\rho_3\rho_4}^{\mu,\mu'}(q,q')
&=&
\sum_{l,l',m,m',\sigma_i} U_{\mu\rho_1;l\sigma_1}({\bm q})
U_{\mu\rho_2;l'\sigma_2}(-{\bm q})
\nonumber \\ && \times 
U_{\mu\rho_3;m\sigma_3}({\bm q}')^* U_{\mu\rho_4;m'\sigma_4}(-{\bm q}')^*
\nonumber \\ && \times
V_{\sigma_1\sigma_3;\sigma_4\sigma_2}^{l,m;m',l'}(q-q'),
\end{eqnarray}
where $U_{\mu\rho;l\sigma}({\bm k})=\langle \mu\rho,{\bm k}| l\sigma,{\bm k} \rangle$
is the unitary matrix ($l$ is the $d$-orbital, $\sigma$ is the real spin).
We approximate that ${\hat V}_{\sigma_1\sigma_2;\sigma_3\sigma_4}(q)$
is given by the SC-VC$_\Sigma$ method without the SOI,
introduced in Ref.\ \cite{Onari-Hdoped}.
(Note that ${\hat V}_{\sigma_1\sigma_2;\sigma_3\sigma_4}(q)
=-({\hat V}^c(q)\delta_{\sigma_1,\sigma_2}\delta_{\sigma_4,\sigma_3}
+{\hat V}^s(q){\vec \sigma}_{\sigma_1,\sigma_2}\cdot{\vec \sigma}_{\sigma_4,\sigma_3})/2$,
where ${\hat V}^{c(s)}$ is the interaction for the 
charge (spin) channel with $\Delta$-VC \cite{Onari-Hdoped}.)
This approximation is reasonable since the pairing interaction 
is induced by higher-energy processes, whereas
the SOI is not important except near the band crossing points.
In fact, the relation $\chi^s_z({\bm q})\approx \chi^s_{x(y)}({\bm q})$ is 
obtained in the RPA for $\lambda=0.4$ even for 
the spin Stoner factor $\alpha_S=0.99$.
Thus, spin-fluctuation-driven VC will give strong orbital fluctuations
even in the presence of the SOI.

The six-fold degeneracy of the TSC state is lifted by the SOI.
For $U=3.05$ and $J/U=0.12$ at $T=0.05$,
the two-dimensional $E_u$ TSC state 
($d_x=d_y=0$, $d_z=\Delta_x,\Delta_y$) is obtained
with the largest eigenvalue $\lambda^E=0.237$.
($\lambda^E=0.234$ for the $A_{1g}$ SSC state, and 
$\lambda^E=0.232$ for the helical $A_{1u}$ TSC state;
$d_x=\Delta_x$, $d_y=\Delta_y$, $d_z=0$.)
The $E_u$ TSC gap functions at the lowest Matsubara frequency 
are shown in Figs.\ \ref{fig:AP-SOI} (d)-(f)
in the presence of the SOI with $\lambda=0.4$.

In the case of $E_u$ TSC state, the chiral superconductivity
$d_z=\Delta_x+i\Delta_y$
is expected to be realized below $T_{\rm c}$.
Figure \ref{fig:AP-SOI} (g) shows the 
magnitude of the chiral gap state
$|\Delta^\delta({\bm k})|=\sqrt{\Delta_x^\delta({\bm k})^2+\Delta_y^\delta({\bm k})^2}$
for $\delta=\alpha,\beta,\gamma$.
Therefore, relatively large gap on the FS$\gamma$ is
induced by that on the FS$\beta$ around $\theta\sim\pi/4$ due to the SOI,
which is approximately given as
$|\Delta^\gamma({\bm k})| \sim r_{xz+yz}({\bm k})|\Delta^\beta({\bm k})|$.
This result is consistent with the experimental report
by the field-orientation dependent specific heat measurement 
\cite{Deguchi}.
It is our important future problem to explain the approximate 
$T$-linear specific heat.
Also, we will perform the RG+cRPA analysis 
for the three orbital Hubbard model with SOI in future.

In contrast, if the TSC is mainly realized on the FS$\gamma$,
the gap on the FS$\alpha$ induced by the SOI
should be very small, since the orbital mixing 
is very small in the $\alpha$-band.
In this case, the ``residual specific heat'' appears
for $T\ll T_c$ due to the tiny gap on the FS$\alpha$,
although it is inconsistent with the specific heat measurement.

To summarize this subsection,
it is found that substantial superconducting gap on FS$\gamma$ is induced 
from the FS$\beta$ due to the large SOI of Ru-atom (proximity effect),
since SOI-induced orbital-mixture is very large between FS$\beta$ and FS$\gamma$.
We also briefly discussed 
the ${\bm d}$-vector \cite{Ng,Yanase,SOPT}
in the presence of the SOI,
which is closely related to the important topological properties of 
the TSC state
\cite{Read,Matsumoto,Furusaki,Tanaka-Rev}.

\vspace*{-.2cm}

\section{Summary}
\label{sec:summary}

\vspace*{-.2cm}

In the present paper,
we proposed that the orbital+spin fluctuation-mediated TSC 
is realized in Sr$_2$RuO$_4$ 
by applying the RG+cRPA method and the SC-VC$_\Sigma$ method
to the multiorbital Hubbard models.
Thanks to the VC that is neglected in the RPA,
strong orbital and spin fluctuations at ${\bm q}\sim{\bm Q}$ emerge in the q1D FSs.
Then, the TSC is easily induced by 
the coexisting orbital and spin fluctuations.
The present theory naturally explains
why the TSC overwhelms the SSC in Sr$_2$RuO$_4$ despite the 
strong AFM fluctuations.

We also analyzed the effect of the SOI on the superconducting gap function.
It is found that the large gap on the FS$\gamma$ is induced 
from that on the FS$\beta$ due to the SOI-induced ``orbital-mixture'' 
between FS$\beta$ and FS$\gamma$.
This effect will be important to understand the experimental 
approximate $T$-linear behavior of $C/T$ below $T_{\rm c}$:
If the TSC were mainly realized on the FS$\gamma$,
the ``residual specific heat'' should appear for $T\ll T_c$ since 
the SOI-induced orbital-mixture between FS$\alpha$ and FS$\gamma$ is very small.
It is our important future problem to obtain the phase diagram of 
the $\bm d$-vector in the TSC state.

\textit{Note added in proof.}
Recently, we became aware of the resonant inelastic X-ray scattering
study 
\cite{Fatuzzo},
in which the SOI-induced large orbital mixture in Sr$_2$RuO$_4$ 
discussed in the
present paper had been confirmed experimentally.

\vspace*{.5cm}
\acknowledgements
We are grateful to K.\ Yamada, Y.\ Maeno, Y.\ Matsuda, K.\ Ishida, 
T.\ Takimoto, T.\ Nomura, C.\ Platt, and C.\ Honerkamp for fruitful discussions.
This study has been supported by Grants-in-Aid for Scientific 
Research from Japan Society for the Promotion of Science.
Part of numerical calculations were
performed on the Yukawa Institute Computer Facility.



\begin{thebibliography}{99}
\bibitem{comment-d}
By applying the present method to the single-band square-lattice Hubbard model,
the $d_{x^2-y^2}$-wave SSC gap function is well reproduced
near the half-filling.
\end{thebibliography}
\end{document}